# Interplay of ethaline and water dynamics in a hydrated eutectic solvent: Deuteron and oxygen magnetic resonance studies of aqueous ethaline

Yannik Hinz and Roland Böhmer
*Fakultät Physik, Technische Universität Dortmund, D-44221 Dortmund, Germany*

The impact of hydration on the reorientational dynamics of the deep eutectic solvent (DES) ethaline, a 2:1 molar mixture of ethylene glycol and choline chloride was studied. Its overall response was explored by means of shear mechanical rheology. To achieve component-selective insights into the dynamics of this material, isotope-edited deuteron and oxygen spin-lattice and spin-spin relaxometry as well as stimulated-echo spectroscopy were applied and yielded motional correlation times from above room temperature down to the highly viscous regime. For all temperatures the cholinium anion was found to reorient two times faster than ethylene glycol, while the water and the ethylene glycol molecules display very similar mobilities. While hydration enhances the component dynamics with respect to that of dry ethaline, the present findings reveal that it does not detectably increase the heterogeneity of the solvent. Merely, the timescale similarity that is found for the hydrogen bond donor and the water molecules over a particularly wide temperature range impressively attests to the stability of the native solvent structure in the "water-in-DES" regime.

## I. Introduction

A countless number of different ingredients enable one to produce eutectic solvents and thus to tailor liquids with tunable material properties. An ever increasing list of practical applications in the fields of, e.g., battery technology, electrodeposition, carbon dioxide sequestration, catalysis, and pharmaceutical research.[1,2] Although there may be different views on how to define deep eutectic solvents (DESs),[3,4] these liquids are usually formed by a hydrogen bond acceptor (HBA), e.g., choline chloride (ChCl) which comprises the cholinium cation and the chloride anion, mixed with a suitable hydrogen bond donor (HBD), e.g., ethylene glycol (EG). The resulting 2EG:1ChCl mixture, is usually called ethaline. Although it is neither "deep" with respect to the melting points of the pure components, nor at the eutectic composition,[5] together with the 2:1 mixtures of glycerol or urea and ChCl, ethaline is in fact among the most widely studied DESs.[6] It was variously emphasized that in terms of structural parameters there is nothing "magic" regarding the stoichiometric (here 2:1) molar mixing ratio.[4,7,8] For the EG:ChCl system the eutectic composition occurs in fact at a molar ratio of about 4.8:1.[5,9,10] Interestingly, for this molar ratio the rotational dynamics as well as the charge transport are enhanced with respect to the 2:1 composition.[10]

With plenty of hydrogen bonds and ionic species present in these mixtures, it is no surprise that they are highly hygroscopic. Hence, unless special precautions are taken, some water is always present in these systems. However, as has variously been pointed out, for practical applications this contamination can even be advantageous.[11,12] Thus, depending on the fraction of added $H_2O$, which can act as a HBD *and* as a HBA, water incorporation may pave the way to novel applications such as aqueous supercapacitors[13] and lead to all kinds of interesting behaviors.[14,15,16,17,18,19,20]

Based on molecular dynamics simulations of slightly moist (<5 wt%) EG:ChCl it was reported that the ionic components, $Ch^+$ and $Cl^-$, prefer to form H-bonds with water rather than with EG.[21,22] This preference does not seem to impart the diffusion of the ionic species,[23] an effect which slightly favors the association among the EG molecules.[17] Thus, with increasing water content, the size of the ethaline associates is expected to decrease somewhat, while water-water proximities increase without significantly altering the gross features of the DES.[24] This relatively smooth character of the water incorporation into the DES matrix, more precisely the sequestration of water around the cholinium cations[25] has been held responsible for the observation that in ethaline and related systems this "water-in-DES" structure persists to a water fraction of about 30 wt%.[26,27] At higher concentrations the native DES structure becomes more and more disrupted, leading first to a "DES-in-water" scenario and eventually to the formation of aqueous electrolyte solutions.[21,28]

Not only the DES structure but also its dynamics is prone to undergoing changes as water is added to DESs. For the EG:ChCl:$H_2O$ mixtures, on which the present study centers, transport quantities such as the translational diffusion[29] and the viscosity were examined near and above ambient temperatures.[11,30] For water concentrations of at least up to about 30 wt%, the overall fluidity as well as the diffusivities of the $Ch^+$ and the EG moieties increase with respect to those of anhydrous DES, where the light and uncharged EG diffuses faster than $Ch^+$.[29] A study of the water dynamics separate from those of the other components is precluded in the presence of the exchange processes taking place among the oxygen bonded protons of $Ch^+$ and of EG with those of $H_2O$.[11,29]

Many DESs are glass forming so that their dynamics can be studied over wide temperature ranges down to the glass transition which for dry ethaline has been reported to occur at $T_g \approx 158$ K.[31] This advantageous situation not only applies for anhydrous but also for wet ethaline up to about 30 wt% of water where $T_g$ is about 9 K lower.[32] Upon sufficient cooling, ice formation drives a phase separation of the ethaline and water components For larger concentrations.[32] Yet, at higher temperatures, in the water-rich single-phase regime of the phase diagram, in a way as a pre-cursor phenomenon, a significant decoupling of charge transport (as determined using conductivity measurements) and molecular reorientation (as detected using dielectric spectroscopy)[33] is observed for >30 wt% of water, but not below this threshold. In dry DESs, the question of a so-called rotation–translation decoupling was also addressed by exploiting dielectric in conjunction with mechanical spectroscopy.[34]

It has been reported that an increase of water (from 1 to 30 wt%) that is added to ethaline slows the structural relaxation by about a factor of 30 (0.21 ms / 7.1 μs at 180 K) and leads to an enhanced dielectric relaxation strength.[12] The latter effect was attributed to an emerging prevalence of EG−water interactions, "consistent with the subsequent increase in the effective dipole moment."[12] In other words, with this kind of coupling present, in hydrated DESs a separate study of the water dynamics again turns out to be difficult in most experimental studies.

To obtain insights specifically into the dynamics of water, dedicated measures are obviously required. Owing to the presence of the mentioned proton exchange effects, the hydrogen-deuteron isotope substitution technique that was successfully employed in component-selective neutron[18,35,36,37,38] and nuclear magnetic resonance (NMR) studies[39,40] of dry DESs, is not suitable for a selective investigation of the water dynamics.

To accomplish the goal of tracking the water dynamics in a proton exchanging system nevertheless, the use of $^{17}$O labeled water suggests itself as a suitable means. Therefore, the present work exploits this approach as well. That is, we employ oxygen NMR to monitor the dynamics of $H_2^{17}O$ that is added to selectively component-deuterated ethaline. This kind of spin tagging allows us to examine the EG, the Ch$^+$, and the water moieties in an isotope edited fashion. Furthermore, with the goal to study the component-averaged low-temperature molecular hopping or transport events (as different from the molecular reorientations that have variously been studied in hydrated ethaline using dielectric spectroscopy),[11,33] in addition to NMR, in the present work we also utilize shear rheology.

## II. Experimental Methods and Details

Like in previous work, here we used ethylene glycol-d$_4$ (EG-d$_4$, chemical purity 99%, isotopic enrichment 98.9 atom%, Santa Cruz Biotechnology), choline chloride-d$_4$ (ChCl-d$_4$, purity 98%, enrichment ≥ 98%, Sigma Aldrich/EQ Laboratories), unlabeled EG (purity 99.8%, Sigma Aldrich), as well as unlabeled ChCl (purity ≥ 98%, Alfa Aesar). Furthermore, $H_2^{17}O$ from Sigma Aldrich, listed with an isotopic enrichment of 40-44.9 atom% and a chemical purity of 99%, was employed to prepare the NMR samples. First, specifically isotope labeled ethaline was prepared as described elsewhere[39,40] and then appropriate amounts of water were added.

The molar fractions of the unlabeled mixtures (used for some of our dielectric spectroscopy and rheology experiments) are expressed as 2EG:1ChCl:$n$H$_2$O with the molar water-to-ethaline (or -choline) ratio given by $n$ = 2.9. For the labeled 2EG:1ChCl-d$_4$:$n$H$_2^{17}$O and 2EG-d$_4$:1ChCl:$n$H$_2^{17}$O samples, $n$ is 2.8 and 2.9, respectively. In all cases this corresponds to 17 wt% of water. For brevity of notation, the variables (1, 2, and $n$) that characterize the molar compositions are omitted in the following.

The rheological measurements were carried out in a plate-plate geometry using an Anton-Paar MCR 502 rheometer. A detailed description of the setup can be found elsewhere.[34,41] Since hydrated ethaline was already thoroughly studied by means of dielectric spectroscopy,[11,33] in the present work this method was used only to ensure compatibility with the literature data and to further check for possible H/D isotope effects. In other DESs such effects were already addressed and found insignificant.[18,35,36,39,40] To examine whether an $^{16}$O/$^{17}$O effect is detectable in hydrated ethaline, as supplementary material we present dielectric measurements on suitably labeled samples. These results demonstrate the absence of an $^{16}$O/$^{17}$O isotope effect, a plausible finding since the same situation applies even for the low-temperature dynamics of pure water.[42]

All NMR samples were hermetically sealed in glass tubes after subjecting them to a freeze and pump cycle. The $^2$H NMR experiments were conducted using a home-built spectrometer at a Larmor frequency of $\omega_D = 2\pi\times45.6$ MHz. The oxygen NMR experiments were carried by means of a Bruker Avance III spectrometer at $\omega_O = 2\pi\times54.3$ MHz (and $\omega_H = 2\pi\times400.1$ MHz) in an $^{17}$O/$^1$H double-resonance probe head manufactured by NMR Service Co. Continuous-wave proton decoupling, typically at a power level of 35 to 50 W, was employed to avoid an undesired impact of the water protons on the oxygen magnetization. Proton decoupling turned out necessary for temperatures $T$ < 180 K when acquiring absorption spectra and during the evolution times of the stimulated-echo measurements. All oxygen-17 spectra are referenced with respect to liquid $^1$H$_2^{17}$O at 291 K. The π/2 radio-frequency pulses were typically 2.5…3.5 μs long for $^2$H. For oxygen they were 9.7...11.2 μs long when the entire $^{17}$O spectrum (satellite plus central transitions) was irradiated and 2.4...2.6 μs at lower temperatures when only the central transition was excited.

To determine the spin-lattice relaxation times $T_1$ we measured the longitudinal magnetization recovery $M_z(t)$ using saturation or inversion recovery techniques. Then, $M_z(t)$ was fitted using the stretched exponential function

$$M_z(t) \propto \exp\left[-(t/T_1)^{\mu_1}\right]. \quad (1)$$

Likewise, for the measurement of the spin-spin relaxation times $T_2$ via the transverse magnetization decay $M_{xy}(t)$, we usually applied Hahn-echo and solid-echo pulse sequences for the $^{17}$O and the $^2$H experiments, respectively. The resulting $M_{xy}(t)$ curves were fitted using

$$M_{xy}(t) \propto \exp\left[-(2t/T_2)^{\mu_2}\right]. \quad (2)$$

For temperatures above about 190 K we find that $M_z(t)$ and $M_{xy}(t)$ are both exponential ($\mu_1 = \mu_2 = 1$) within experimental uncertainty as is in fact expected in the regime of fast motions for deuterons[43] as well as for halfinteger quadrupolar spins such as $^{17}$O.[44,45,46]

## III. Results and Analysis

### A. Overall dynamics studied by mechanical spectroscopy

In Fig. 1 we show the real and imaginary part of the complex shear mechanical modulus, $G^* = G' + iG''$, of EG:ChCl:H$_2$O for a range of low temperatures. The shear storage component is seen to display well defined steps accompanied by peaks in the shear loss. Both features move through the frequency window as the temperature is altered. From the loss peak maxima one may read out characteristic time constants. However, in order to facilitate the comparison with relaxation times from dielectric spectroscopy, it is more appropriate to analyze the mechanical data in terms of the susceptibility format, in other words, in terms of the complex shear compliance, $J^* = 1/G^*$, as we have emphasized in previous DES studies.[34,41] Recognizing that the loss peaks displayed in Fig. 1(b) are broader than expected in the presence of only a single relaxation time, for the analysis of the associated compliance we use the Cole-Davidson expression[47]

$$J^*(\nu) = J_\infty + \frac{\Delta J}{(1 + 2\pi i\nu\tau_J)^{\gamma_J}} + \frac{1}{2\pi i\nu\eta_0}. \quad (3)$$

Here, the exponent $0 < \gamma_J \leq 1$ characterizes the underlying distribution of the characteristic (retardation) time constants, $\tau_J$, and $\Delta J = J_s - J_\infty$ is the retardation strength, i.e., the difference between the (recoverable) steady-state ($\nu \to 0$) compliance, $J_s$, and the high-frequency compliance, $J_\infty$. Analogous to typical dielectric analyses (where often electrical conductivity effects need to be taken into account), Eq. (3) includes a flow term. In the mechanical spectroscopy of liquids, the role of the dc conductivity is played by the zero-shear-rate fluidity, $1/\eta_0$. Exploiting the Maxwell relation $1/\eta_0 = J_\infty/\tau$, one recognizes that for the present analysis the zero-shear viscosity $\eta_0$ needs not be known explicitly.

We fitted the experimental data using $G^* = 1/J^*$ with $J^*$ given by Eq. (3). In Fig. 1 the results are represented as solid lines and are seen to work very well. As parameters (including their estimated uncertainties) we find $J_\infty = 2.2 \pm 0.2$ GPa$^{-1}$, $\Delta J = 8.5 \pm 0.8$ GPa$^{-1}$, and $\gamma_J = 0.37 \pm 0.04$ as well as the time constants given in Sec IV B.

It should be noted that the present measurements are particularly useful since, owing to an enhanced crystallization tendency, frequency-resolved rheology could not be carried out for dry ethaline.[34]

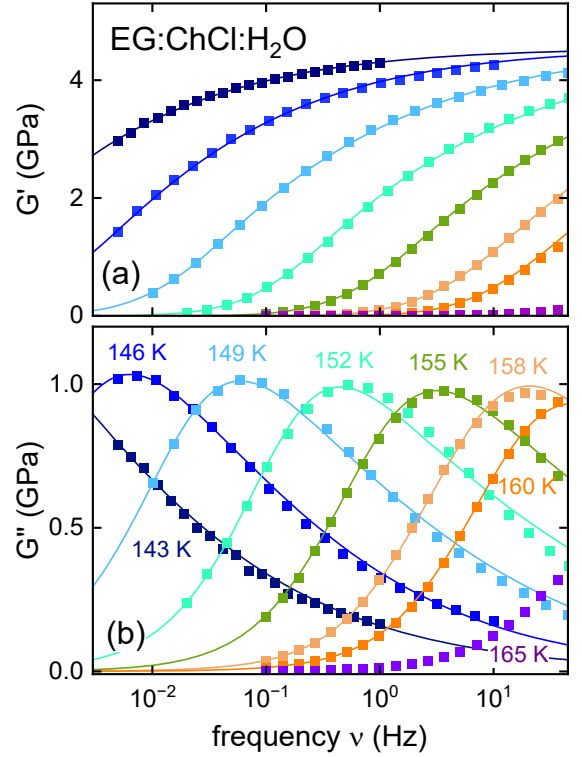

Fig. 1. (a) Storage part $G'$ and (b) loss part $G''$ of the complex shear mechanical modulus of EG:ChCl:H$_2$O are depicted as symbols. The solid lines represent fits using Eq. (3).

### B. Ethaline and water contributions selectively studied by deuteron and by oxygen NMR

Using NMR, the reorientational dynamics is often accessed using absorption line shapes, via spin relaxation measurements, and by means of stimulated-echo spectroscopy. Results obtained using all of these three techniques are presented and discussed in Secs III B 1, 2, and 3, respectively.

#### 1. Quadrupolar absorption spectra

One of the reasons to determine NMR spectra is that they are needed in quantitative analyses of spin-relaxation experiments. As quadrupolar probes, the resonances of the $^2$H nucleus (spin quantum number $I = 1$) and of the $^{17}$O species ($I = 5/2$) are governed by the quadrupolar coupling constant $C_Q = eqeQ/h$ (in units of Hertz) and the asymmetry parameter $\eta = (V_{XX} - V_{YY})/V_{ZZ}$.[48] Here, $|eq| = |V_{ZZ}| \geq |V_{YY}| \geq |V_{XX}|$ designate the Cartesian components of the electrical field gradient (EFG) tensor in its principal axis system. Let the angles $\theta$ and $\phi$ define the orientation of the EFG tensor that characterizes the molecular C-D bonds with respect to the external magnetic field in the usual fashion. Then, the angular dependence of the quadrupolar frequency (derived using first-order perturbation theory) is given by[49]

$$\omega_D = \pm\tfrac{1}{2}\delta_D\left(3\cos^2\theta - 1 - \eta_D\sin^2\theta\cos 2\phi\right). \quad (4)$$

Here, $\delta_D = \frac{3}{4} e^2 qQ / \hbar$ is referred to as the quadrupolar anisotropy parameter and the ± signs correspond to the nuclear $-1 \leftrightarrow 0$ and $0 \leftrightarrow +1$ transitions. We point out that for $\eta = 0$, Eq. (4) reduces to $\omega_D = \pm \delta_D P_\ell(3\cos\theta)$ where the rank of the Legendre polynomial is $\ell = 2$.

As supplementary material we present a low-temperature deuteron spectrum for EG-d$_4$:ChCl:H$_2^{17}$O where the separation of the outer edge singularities yields $2\delta_D$ and the asymmetry parameter can be assessed from the separation, $\Delta\nu = (1 - \eta_D)\delta_D/2\pi$, of the horn singularities.[50] In harmony with previous work[39,40] for both, the EG and the Ch deuterated species we find $\delta_D = 2\pi \times 125 \pm 2$ kHz and $\eta_D \approx 0$. The $^2$H spectra can be fully excited using the relatively short radio-frequency pulses that are employed in the present study.

By contrast, for $^{17}$O a nonselective excitation of the spectrum as a whole, which comprises central and satellite transitions, is possible only at high temperatures. Here, in the regime of full motional narrowing the resonances of all allowed nuclear transitions overlap and the overall absorption line is relatively narrow, see Fig. 2. Upon cooling the spectra successively broaden until about 230 K.

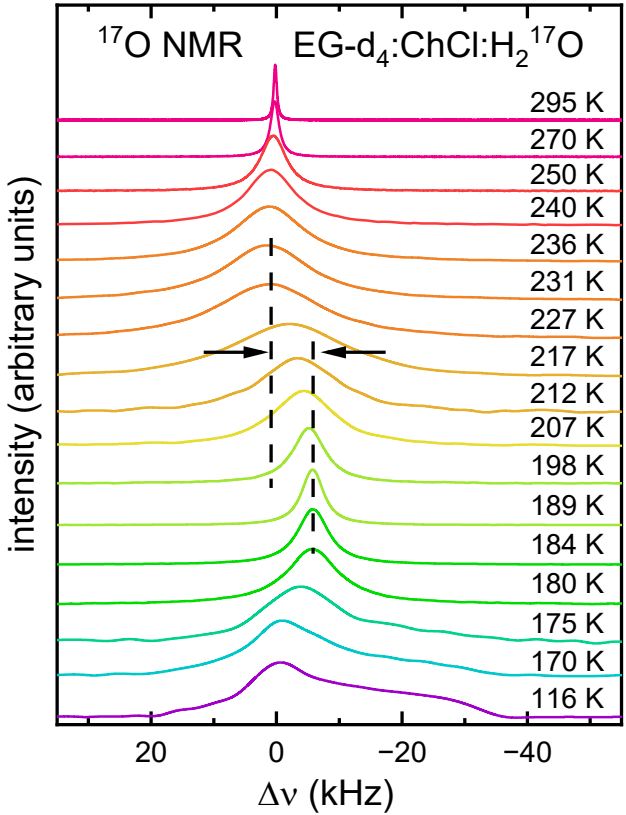

Fig. 2. (a) Experimentally determined $^{17}$O spectra recorded for EG-d$_4$:ChCl:H$_2^{17}$O at a pulse separation of 25 μs. The spectra are successively vertically shifted for visual clarity and normalized to their maximum intensity. The double arrow marks the magnitude of the observed frequency shift which is 6.7 ± 0.5 kHz.

At lower temperatures one is left with spectra originating from only the central $-1/2 \leftrightarrow +1/2$ transition of the $^{17}$O nucleus. The orientation dependence of the corresponding second-order central-transition quadrupolar frequency is given by[51,52]

$$\omega_O = \xi[A(\phi,\eta)\cos^4\theta + B(\phi,\eta)\cos^2\theta + C(\phi,\eta)]. \quad (5)$$

This expression is thus more complex than for the first-order quadrupolar interactions, cf. Eq. (4), and can be viewed to involve Legendre polynomials of rank $\ell = 4, 2,$ and $0$.[53] In Eq. (5), the quadrupolar coupling is encoded in terms of

$$\xi = \frac{3}{2}\frac{\Omega_Q^2}{\omega_O}\left(I(I+1) - \frac{3}{4}\right). \quad (6)$$

with $\Omega_Q = 2\pi C_Q / [2I(2I-1)]$. The azimuthal angle dependent factors $A$, $B$, and $C$ that appear in Eq. (5) can be written as

$$A(\phi,\eta) = -\frac{27}{8} + \frac{9}{4}\eta\cos(2\phi) - \frac{3}{8}[\eta\cos(2\phi)]^2$$
$$B(\phi,\eta) = \frac{30}{8} - \frac{1}{2}\eta^2 - 2\eta\cos(2\phi) + \frac{3}{4}[\eta\cos(2\phi)]^2 \quad (7)$$
$$C(\phi,\eta) = -\frac{3}{8} + \frac{1}{3}\eta^2 - \frac{1}{4}\eta\cos(2\phi) - \frac{3}{8}[\eta\cos(2\phi)]^2.$$

It turns out that the quadrupolar interaction dominates the $^{17}$O resonance of the H$_2$O molecule by far[54] and other interactions, except for the chemical shielding anisotropy (CSA), will here be neglected. According to the Haeberlen convention[55] the CSA is defined as $\Delta\sigma = \sigma_{zz} - \frac{1}{2}(\sigma_{xx} + \sigma_{yy})$, where $\sigma_{ii}$ denotes the eigenvalues of the CSA tensor in its principal axis system. For liquid water, depending on the hydrogen-bond environment, $\Delta\sigma$ was estimated to be in the range from 30 to 40 ppm,[54] close to values found for H$_2^{17}$O in various crystalline solids.[56,57,58,59] In the 9.4 T field that is used for the present $^{17}$O NMR experiments, this corresponds to about 1.6 to 2.1 kHz.

The supplementary material demonstrates that the spectral shape is similar but slightly broader as compared to that reported for $^{17}$O in (stack-disordered) crystalline and (low- and high-density) amorphous ice.[60] Thus, the quadrupolar coupling constant $C_Q = 6.66$ MHz and the asymmetry parameter $\eta = 0.935$ that were reported for hexagonal ice,[61] roughly apply also for EG-d$_4$:ChCl:H$_2^{17}$O. From spin relaxation experiments on *liquid* water quadrupolar coupling constants reaching 7.9 to 8.3 MHz were reported and asymmetry parameters $\eta$ as low as 0.75 were found.[62,63] Here, one should note that smaller $\eta$ reflect smaller H-O-H angles.[64] The shape of the $^{17}$O spectra depends on the relative orientation of the EFG and CSA tensors. Even for crystals,[58] it is often described by the set of Euler angles $(\alpha,\beta,\gamma)$ close to $(90°,90°,0°)$ that are expected from the C$_{2v}$ symmetry of the isolated water molecule. Therefore, for the spectral calculations in the supplementary material we use the listed angles.

Another salient feature of the spectra shown in Fig. 2 is the second-order dynamic frequency shift that is given by[65]

$$\nu_{CG,Q} = -\frac{3}{40}\underbrace{\frac{I(I+1)-\frac{3}{4}}{I^2(2I-1)^2}}_{=3/500\ \text{for}\ I=5/2}\frac{C_Q^2(1+\frac{1}{3}\eta_O^2)}{\nu_O}. \quad (8)$$

This expression shows that in fact it is the quadrupolar product, $P \equiv C_Q(1+\frac{1}{3}\eta_O^2)^{1/2}$, that determines the magnitude of the dynamic frequency shift. The above quadrupolar parameters yield $P_{ice} \approx 7.6$ MHz for solid ice and $P_{water} \approx 8.7$ MHz for liquid water. From Eq. (8) and the shift marked by the double arrow in Fig. 2, the quadrupolar product is estimated to be $7.8 \pm 0.3$ MHz.

Another interesting feature of the spectra shown in Fig. 2 is the non-monotonic temperature dependence of the experimentally determined overall full linewidth at half maximum, $\Delta\nu_{1/2}^{exp}(T)$. This finding is comprehensively discussed in the terms of the associated transverse spin relaxation time $T_2^* = 1/(\pi\Delta\nu_{1/2}^{exp})$ to which we turn next, in Sec III B 2.

### 2. Reorientations probed by spin relaxation

To probe the component selective dynamics in hydrated ethaline, we determined spin relaxation times over wide temperature ranges. Fig. 3 compiles the $T_1$ and $T_2$ times collected in the course of the present work.[66] These include oxygen and deuteron NMR measurements for each of the doubly labeled EG-$d_4$:ChCl:H$_2^{17}$O and EG:ChCl-$d_4$:H$_2^{17}$O mixtures. Most aspects of the present data are reminiscent of the general patterns known from the spin relaxation of other glass forming liquids. Therefore, the assessment of the temperature dependent spin relaxation for the $I = 1$ and the $I > 1$ nuclei that is required for a detailed understanding of the present results closely follows the analyses carried out for other supercooled liquids.[67,68,69] Focusing first on the oxygen resonance, it may suffice to briefly recapitulate the most characteristic features.

Using the spectral density $J(\omega)$ which in the simplest case of a single correlation time $\tau_c$ obeys the Bloembergen-Purcell-Pound (BPP)[70] form, $J_{BPP}(\omega) = \tau_c/[1+(\omega\tau_c)^2]$, the spin-lattice relaxation time is given by[59,71]

$$\frac{1}{T_1} = 2K_O[J(\omega_O) + 4J(2\omega_O)]. \quad (9)$$

The prefactor[72]

$$K_O = \frac{12\pi^2}{400}\underbrace{\frac{2I+3}{I^2(2I-1)}}_{=6\pi^2/625\ \text{for}\ I=5/2}\Delta C_Q^2(1+\tfrac{1}{3}\eta_O^2), \quad (10)$$

does not necessarily reflect the low-temperature value of the quadrupolar product but only its fluctuating part.

The discussion of the spin-spin relaxation is somewhat more involved since various dynamic regimes have to be distinguished. For fast and intermediate motions, $\omega\tau_c \lesssim 1.5$, one has[73]

$$\frac{1}{T_2} = K_O[3J(0) + 5J(\omega_O) + 2J(2\omega_O)]. \quad (11)$$

For $\omega\tau_c \ll 1$, where $T_1 = T_2$, this reduces to

$$\frac{1}{T_2} = \frac{12}{125}\pi^2\Delta C_Q^2(1+\tfrac{1}{3}\eta_O^2)J(0). \quad (12)$$

Using a Cole-Davidson form for the spectral density,[74] one has $J_{CD}(0) = \gamma_O\tau_c$ where $\gamma_O$ designates the appropriate Cole-Davidson exponent. Thus, it is a simple matter to determine the correlation times in the high-temperature regime.

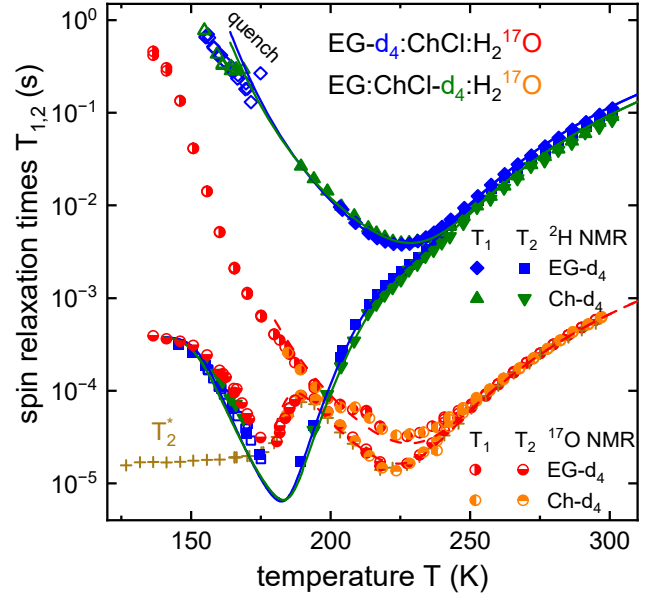

Fig. 3. Temperature dependent spin-lattice and spin-spin relaxation times. Deuteron measurements for EG-$d_4$:ChCl:H$_2^{17}$O (◆,■) and for EG:ChCl-$d_4$:H$_2^{17}$O (▲,▼) are shown. Note that the deuteron data presented for $T < 175$ K were obtained after quenching the samples (open symbols). As described in the text, the solid lines were calculated using Eqs. (9) and (11) in conjunction with Eq. (20). For EG-$d_4$:ChCl:H$_2^{17}$O (◐,◑,+) and for EG:ChCl-$d_4$:H$_2^{17}$O (◐,◑) oxygen-17 measurements from cooling and heating runs are shown. (For brevity, the legend in the figure lists only the distinguishing molecular group, i.e., EG-$d_4$ or ChCl-$d_4$, rather than the entire sample composition). The dashed lines that describe the $^{17}$O data very well. For $T_1$ they were calculated using the equations just mentioned, while for the calculation of the spin-spin relaxation time ($T_2^{theo}$) we employed Eq. (15).

For slower motions, i.e., at somewhat lower temperatures where, as already emphasized, only the central transition is excited, the transverse relaxation rate is governed by the first-order quadrupolar contribution

$$\frac{1}{T_{2c}^{(Q1)}} = 5K_O[J(\omega_L) + J(2\beta\omega_L)], \quad (13)$$

with $\beta = (2/7)^{1/2} \approx 0.53$ for $I = 5/2$.[75] At still lower temperatures (down to about 180 K), where $\tau_c$ does not exceed $\tau_{exc} \approx 20$ μs,[76] also second-order contributions come into play[77,78,79] and the quadrupolar transverse relaxation rate can be approximated as

$$\frac{1}{T_{2c}^{(Q2)}} = \frac{298}{875}\underbrace{\left(\frac{3}{16}\frac{I(I+1)-\tfrac{3}{4}}{I^2(2I-1)^2}\right)^2}_{=(3/200)^2\ \text{for}\ I=5/2}\frac{4\pi^2\Delta C_Q^4(1+\tfrac{1}{3}\eta_O^2)^2}{\nu_O^2}J(0). \quad (14)$$

This means that in the motional regime defined by $\tau_c < \tau_{exc}$, the theoretically expected total width of the absorption spectra should be

$$\Delta \nu_{1/2}^{theo} = \frac{1}{\pi T_2^{theo}} = \frac{1}{\pi T_{2c}^{(Q1)}} + \frac{1}{\pi T_{2c}^{(Q2)}}. \quad (15)$$

A glance at Fig. 3 reveals that in this regime the effective spin-spin relaxation times $T_2^* = 1/(\pi \Delta \nu_{1/2}^{exp})$ computed from the experimentally recorded linewidths $\Delta \nu_{1/2}^{exp}(T)$ indeed agree well with the measured spin-spin relaxation times $T_2$.[80]

For the calculation of the oxygen spin-lattice relaxation times according to Eq. (9) and the spin-spin relaxation times according to Eq. (15), that are shown in Fig. 3 as dashed lines, we employed a Cole-Davidson exponent $\gamma_O = 0.41$ and a quadrupolar product $P_O = 8.5$ MHz. For temperatures below about 180 K, from Fig. 3 it is clear that $T_2^* < T_2$ since the measurement of the latter transverse dephasing time (but not that of the former) involves refocusing.

The two terms contributing to Eq. (15) display opposite ($\propto \tau_c$ or $\propto 1/\tau_c$) dependences, which leads to a minimum linewidth. Hence, if the BPP case applies, a maximum overall $^{17}O$ dephasing time arises when $\tau_{c,T2max} \approx 2.8/[\Delta C_Q(1 + \eta_O^2/3)^{1/2}]$.[72] Otherwise, a numerical analysis of Eq. (15) is mandatory[81] which for the present parameters yields $\tau_{c,T2max} \approx 2.3$ µs. The corresponding time constant as well as those derived from Eq. (12) are discussed in Section IV B.

Before doing so, let us turn to the deuteron spin relaxation times that are also shown in Fig. 3. Owing to the much smaller quadrupolar coupling (for $^2H$ as compared to $^{17}O$), the deuteron spin (-lattice) relaxation times are relatively long. It is also obvious that, within experimental uncertainty, the $^{17}O$ spin relaxation times for the Ch$^+$ deuterated and the EG deuterated species agree while at high temperatures their $^2H$ spin relaxation times differ somewhat. Specifically, the deuteron $T_1$ and $T_2$ times are longer at the EG site which implies that in hydrated ethaline the C-D bonds of EG reorient faster than those of Ch$^+$, a conclusion that was similarly drawn for dry ethaline.[39]

As discussed in Ref. 39 as well, a pronounced crystallization tendency precluded the acquisition of spin relaxation times for anhydrous ethaline in relatively wide temperature ranges. For the present investigation, after slowly cooling the sample to the 165 to 175 K range the $^2H$ data also surfaced some irregularities (see the supplementary material). To circumvent them, subsequent to a rapid cooldown of the sample, additional $^2H$ spin relaxation times were obtained in and below this temperature range that are included in Fig. 3. Together with the slow-cool data they display a smooth overall temperature dependence.

The quantitative analysis of the $^2H$ spin-lattice relaxation times is largely similar to that for $^{17}O$, except that for $I = 1$ (i) the numerical factor appearing in Eqs. (10) and (11) is $K_Q = 3\pi^2/20$, and (ii) $J(0)$ in Eq. (11) is to be replaced by $J(\omega_{Q,\text{eff}})$ with $\omega_{Q,\text{eff}}$ on the order of the quadrupolar anisotropy parameter.[82] Furthermore, to capture the behavior of the deuteron $T_2$ below its minimum, where the molecular time scale typically is $1/\delta_D \approx 1$ µs, it is appropriate to add a dephasing term, $1/T_{2,\text{dip}} = \sigma_{\text{dip}} 2\ln 2$, to Eq. (11).[82] This term accounts for the homonuclear ($^2H$-$^2H$) dipolar dephasing that gains increasing importance for temperatures below the $T_2$ minimum. For the present calculations we use the same dipolar interaction parameter, $\sigma_{\text{dip}} = 2\pi \times 300$ Hz, as for dry ethaline[39] and an effective anisotropy parameter $\omega_{\text{eff}} = 2\pi \times 40$ kHz. Fig. 3 shows that this way a good description of the low-temperature deuteron $T_2$ behavior of hydrated ethaline is achieved.

### 3. Slow motions probed by stimulated echoes

The NMR methods dealt with so far are most sensitive to motions on the scale of microseconds or faster. In order to cover as large a dynamic range as possible, stimulated-echo spectroscopy was additionally carried out. This technique can yield access to reorientational motions taking place in the ultraslow domain (beyond ≈0.1 ms). Emanating from the Jeener-Broekaert sequence,[83] three appropriately phase cycled pulses are applied to generate $^2H$ or $^{17}O$ signals of the form[84]

$$F_2^{\cos}(t_p, t_m) = c_I \langle \cos(\omega(0)t_p) \cos(\omega(t_m)t_p) \rangle, \quad (16a)$$

and

$$F_2^{\sin}(t_p, t_m) = s_I \langle \sin(\omega(0)t_p) \sin(\omega(t_m)t_p) \rangle. \quad (16b)$$

Here, the $\langle ... \rangle$ brackets indicate an ensemble average. These correlation functions are often used to probe the molecular motion on the scale set by the (variable) mixing time $t_m$. In favorable cases the (typically fixed) pulse separation $t_p$ can be exploited to obtain information regarding the motional geometry of the molecular segments under study. Although the signals that appear in Eq. (16) look similar for a wide range of nuclear species[85] one has to realize that for the two isotopes under scrutiny, the excitation conditions differ. For quadrupolar nuclei with halfinteger spin, where only the central transition is excited, an additional signal damping occurs on the same $T_1$ scale for the sine as well as for the cosine experiment. Based on independently measured longitudinal magnetization recoveries, $M_z(t_m)$, these damping effects can easily be taken into account. Furthermore, for $^{17}O$ the amplitude prefactors are $c_{5/2} = s_{5/2} = 3/35$ if signal maximizing radio-frequency pulses are utilized.[86] Incidentally, the currently employed central-transition technique is not only applicable in the context of $^{17}O$ NMR,[53,67,86,87] but was successfully demonstrated for $I = 3/2$ nuclei such as $^{11}B$,[88] $^{35}Cl$,[69] and $^{87}Rb$[89] as well.

For the deuteron ($I = 1$), Eq. (16) assumes that the entire spectrum is excited. Here, the maximum amplitude prefactors are $c_1 = 1$ and $s_1 = 3/4$.[49] The measurement of deuteron sine correlation functions, cf. Eq. (16b), is often hampered by a signal damping that involves a spin relaxation time (called $T_{1Q}$) which for supercooled liquids is usually inaccessible in an independent fashion.[90] Conversely, for deuteron cosine correlation functions, cf. Eq. (16a), the signal damping is simply given $T_1$.

signals $S_2(t_m) = F_2(t_m) M_z(t_m)$ shown in Fig. 4, they were fitted using a stretched exponential function

$$F_2(t_m) \propto \exp[-(t_m/\tau_c)^\beta]. \quad (17)$$

Here, the Kohlrausch exponent $\beta$ quantifies the nonexponentiality of the correlation decay. The resulting time constants will be discussed in Sec IV B. In preliminary fits, the $\beta$ exponent was treated as a free parameter and subsequently it was fixed to a value of 0.46 which was found to provide a good description for all data sets from $^2$H and from $^{17}$O NMR.

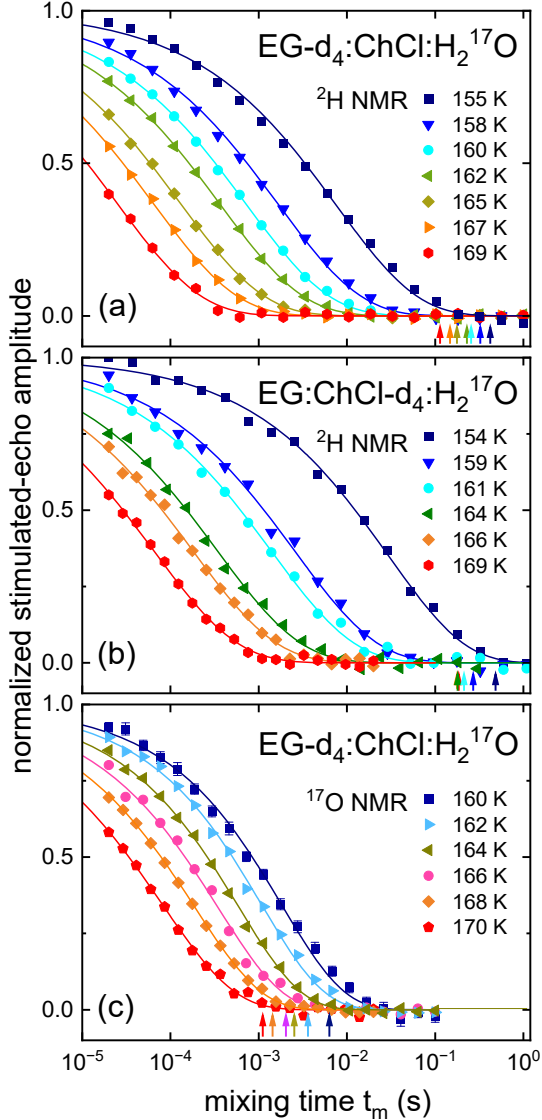

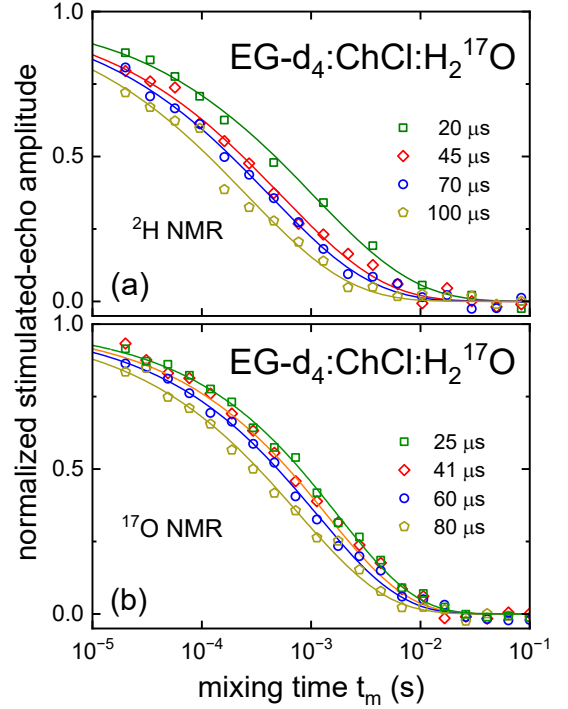

Fig. 4. Normalized cosine stimulated-echo signals $S_2(t_m)$ recorded for a range of temperatures. Deuteron measurements with $t_p$ set to 20 μs are displayed in panels (a) and (b) for EG-d$_4$:ChCl:H$_2^{17}$O and EG:ChCl-d$_4$:H$_2^{17}$O, respectively. Panel (c) features oxygen-17 measurements with $t_p = 25$ μs for EG-d$_4$:ChCl:H$_2^{17}$O. Appropriately colored arrows highlight the 1/e buildup times of the longitudinal magnetization. In panel (c) temperatures below 160 K are not shown because here the mixing time window is limited by spin relaxation. In all panels the solid lines reflect fits using Eq. (17).

Fig. 4(a) and (b) present stimulated deuteron echo signals $S_2(t_m)$ that we measured for the EG and the Ch$^+$ moieties in hydrated ethaline for a range of temperatures. To determine the reorientational correlation time $\tau_c$ from the

Fig. 5. Normalized (a) deuterium and (b) oxygen-17 cos-cos stimulated-echo signals $S_2(t_m)$ of EG-d$_4$:ChCl:H$_2^{17}$O as recorded at 160 K for several evolution times $t_p$. The solid lines reflect fits using Eq. (17).

Fig. 4(c) shows the $^{17}$O results that relate to the slow water motion in EG-d$_4$:ChCl:H$_2^{17}$O. Again we employed Eq. (17) for their quantitative description. Within experimental uncertainty we find that $F_2(t_m)$ decays to zero for all samples and probe nuclei. In particular, this statement implies that jumps among a small number of different sites can be ruled out. Here, one should keep in mind that π flips of the C-D bonds, or for jumps of the H$_2^{17}$O molecules about their C$_{2v}$ axes, would remain undetected. This is because, owing to Eqs. (4) and (5), such jumps leave their respective NMR frequencies invariant.

To enable further tests of the motional geometry, we carried out $^2$H and $^{17}$O stimulated-echo experiments for several evolution times and show the results in Fig. 5(a) and (b), respectively. One observes that the effective decay times depend markedly on $t_p$, unlike what would be expected in the presence of large ($\gtrsim 30°$) jump angles.[91] More detailed consideration including a comparison of the latter $^2$H and $^{17}$O results will be provided next, in Sec IV A.

## IV. Discussion

### A. Comparison of stimulated $^2$H and $^{17}$O echoes

In Fig. 6 we collect the effective timescales $\tau(t_p)$ as on­tained by the fits to the $^2$H and $^{17}$O stimulated-echo curves shown in Fig. 5. First focusing on the deuteron data, one recognizes how the effective time constants become short­er as the evolution times $t_p$ increases. In order to discuss this behavior, it is instructive to compare it with random-walk simulations that describe analogous $^2$H NMR results for other glassformers such as glycerol.[90] Subsequent to adjusting the overall time scale, i.e., $\tau(t_p\to 0)$, the simula­tions are seen to describe the deuteron results for hydrated ethaline well. The reorientational scenario underlying these simulations assumes a bimodal distribution of small-angle (2°) and large-angle (30°) jumps where, in the course of time, the latter occur 50 times less frequent.[90] It has to be realized that in the spirit of Occam's razor this provides probably the simplest parameterization (others have been discussed[92]) of the complex motional geometry character­izing the dynamics of deeply supercooled liquids.

The scenario just described can be viewed as a generali­zation of Anderson's jump model[93] which assumes that molecular reorientation occurs via a succession of jumps about a fixed angle $\alpha_\mathrm{jump}$. In this context it is worthwhile to recall that for oxygen-17, even in the small-$t_p$ limit, Eq. (16) corresponds merely to an $\ell = 4$ correlation function (plus additional terms), while for $^2$H Eq. (16) reflects an $\ell = 2$ correlation function.[94] For an arbitrary rank $\ell$, the Ander­son model predicts a timescale ratio[67,95]

$$r_\ell(\alpha_\mathrm{jump}) \equiv \frac{\tau(t_p \to 0)}{\tau(t_p \to \infty)} = 1 - P_\ell(\cos\alpha_\mathrm{jump}) . \quad (18)$$

In the presence of a distribution of jump angles it has been found that it is the smallest of them that determines $\tau(t_p\to\infty)$.[90] Thus, when comparing an $\ell = 4$ with an $\ell = 2$ response, Eq. (18) predicts that the associated timescales should differ by $r_4(2°)/r_2(2°) \approx 3.3$. This estimate corre­sponds to the limit of small jump angles because the model of rotational Brownian motion predicts an $\ell$-dependent reorientation time of $\tau_\ell = \ell(\ell+1)\tau_c$ so that $\tau_{\ell=2}/\tau_{\ell=4} = 10/3$. However, this ratio should be considered as an upper limit since, as stated below Eq. (6), not only $\ell = 4$ but also $\ell = 2$ terms enter the precession frequencies of the oxygen spins. Thus, the timescale difference between the $^2$H and the $^{17}$O results originating from the discussed effect is $2 \pm 1$ (this is to say it is between $\approx 1$ and $\approx 3$). Thus, if only the effects of an $\ell$ dependence related to Eq. (18) are considered, the net factor for a given $t_p$ is estimated to be $\approx 2.5$ (5 as mentioned in the caption of Fig. 6 divided by $2 \pm 1$).

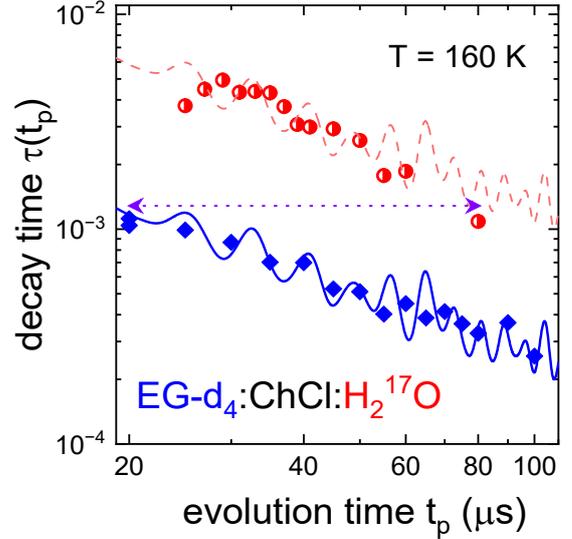

Fig. 6. Stimulated-echo time scales $\tau_\mathrm{EG}(t_p)$ and $\tau_\mathrm{H2O}(t_p)$ measured for EG-d$_4$:ChCl:H$_2$$^{17}$O using $^2$H NMR (diamonds) and $^{17}$O NMR (circles), respectively, as a function of the evolution time at 160 K. The solid lines reflect simulations based on the reorientational model described in the text.[90] The dashed line (which was ob­tained by vertically shifting the solid line by a factor of 5) serves only to illustrate the difference of the oxygen and the deuterium based timescales. The dotted arrow is meant to illustrate the (by a factor of about 4) differing $^2$H and $^{17}$O phase evolutions that are implied by Eq. (19).

However, not only the timescale difference along the y-axis deserves particular attention, but also that along the x-axis. This is because for the discussion of stimulated-echo experiments yet another distinguishing factor exists when comparing, for a given evolution time, the deuterium with the oxygen based timescales. In Eq. (16) the phase argu­ment in the trigonometric function is $\omega t_p$, where the "mean" of the precession frequency $\omega$ may be approximat­ed by the square root of the second moment of the associ­ated frequency distribution (in other words, of the corre­sponding absorption spectrum). Thus, the use of the same $t_p$ for species with differently broad spectra will in general lead to the different phase evolutions. For the considered nuclear species, roughly comparable phase evolutions will emerge in a given time $t_p$ if

$$t_p(^{17}\mathrm{O})\sqrt{M_{2c}(^{17}\mathrm{O})} \approx t_p(^2\mathrm{H})\sqrt{M_2(^2\mathrm{H})} . \quad (19)$$

Here, $M_2(^2\mathrm{H})$ designates the second moment of the $^2$H spectrum[96] and $M_{2c}(^{17}\mathrm{O})$ the (central) second moment of the $^{17}$O central-transition spectrum.[97] Inserting the spectral parameters characterizing the two species in hydrated ethaline, Eq. (19) yields $t_p(^{17}\mathrm{O})/t_p(^2\mathrm{H}) \approx 56$ kHz / 14 kHz.[96,97] Thus, one may say that typically, to achieve com­parable phase evolutions, for the oxygen experiment the evolution time needs to be chosen about 4 times longer than for the deuteron experiment. The vertical arrow in Fig. 6 is meant to illustrate the effective "left shift" of $t_p(^{17}\mathrm{O})$ with respect to $t_p(^2\mathrm{H})$, if only the effect of Eq. (19)



is considered.

The combined effects of Eqs. (18) and (19) imply that despite the different appearances of the measured timescales in Fig. 6, the dynamics of the water molecules are comparable or slightly faster than those of the EG molecules.

## B. Comparison of motional time scales

In Fig. 7 we present the time scales resulting from the stimulated-echo experiments shown in Fig. 4. These data were measured at fixed evolution times. Based on Eq. (17), mean time scales $\langle \tau \rangle = \tau \beta^{-1} \Gamma(1/\beta)$ were extracted, where $\Gamma$ denotes Euler's gamma function. Due to the timing restrictions of the stimulated-echo NMR technique,[61,67] the data cover a relatively narrow temperature range of, here, about 15 K. Thus, in the Arrhenius representation of Fig. 7 the curvature expected in such a plot is barely visible for the NMR data. Merely, the temperature dependence can be well approximated by an effective energy barrier. The corresponding straight lines that are shown in Fig. 7 are all parallel to each other. This observation implies that within experimental uncertainty well-defined timescale ratios exist with respect to the various components that form hydrated ethaline. In particular, for $\tau_{EG}/\tau_{H2O}$ we find a ratio of $4.8 \pm 0.5$ which is compatible with that inferred from Fig. 6. As discussed in Sec IV A, taking into account the effects associated with Eqs. (18) and (19), the water molecules move on a timescale that is about the same or slightly smaller than that characterizing EG.

The comparison of the time scales related to EG and to cholinium is much simpler, because both are monitored using the same nuclear probe (relating to the same $\ell = 2$) at the same $t_p$. From Fig. 7 the $\tau_{EG}/\tau_{Ch}$ ratio turns out to be $0.47 \pm 0.05$, i.e., the bulkier Ch moiety moves somewhat slower than the smaller EG molecule.

To put the stimulated-echo data into perspective, Fig. 7 includes also other times scales for hydrated ethaline. For the dielectric data from Ref. 33 the water-to-ethaline ratios differ from that for the present sample. To account for the concentration effect (slightly larger water contents diminish the relaxation times), on the logarithmic time axis we interpolated the dielectric time scales[33] for $n = 1.7$ (10 wt%) and $n = 3.6$ (20 wt%) to a value of $n = 2.9$ (17 wt%). In Fig. 7 the resulting slow-down factor of from 2 at 173 K to 3 at 158 K (with respect to the 20 wt% sample) is indicated by the vertical arrows.

From Fig. 7 one also notes that the NMR ($\ell = 2$) time scales,[94] for the deuterons) are shorter than the dielectric ($\ell = 1$) time scales. We attribute this time scale difference to effects implied by Eq. (18) from which $\tau_{\ell=1}/\tau_{\ell=2} = 1...3$ follows,[67,95,98] combined with the fact that typically $\tau_{\ell=2}(t_p \to 0) >> \tau_{\ell=2}(t_p)$ with $t_p$ in the 20...25 µs range.[67,90,92]

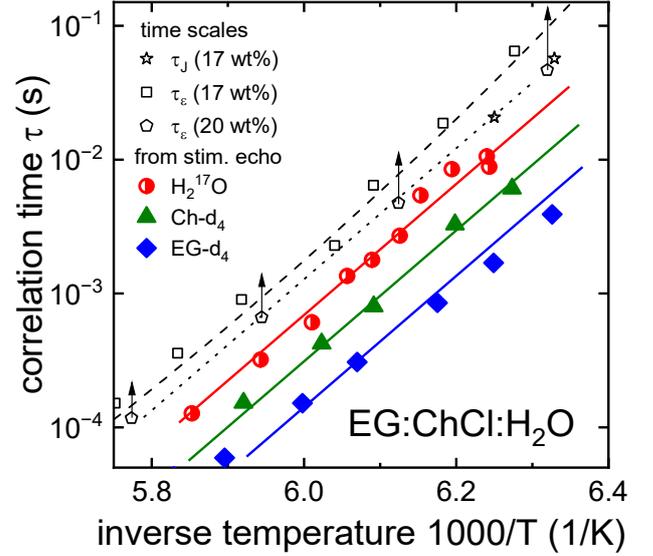

Fig. 7. Arrhenius plot featuring mean time scales from the shear compliance ($\tau_J$, this work), from dielectric spectroscopy ($\tau_\varepsilon$, with 20 wt% H$_2$O by Jani et al.[33] and 17 wt% H$_2$O from this work), and from the stimulated-echo results shown in Fig. 4 for the listed components of hydrated ethaline. As described in the text, the upward pointing arrows are meant to reflect the concentration effect on the dielectric time scales from Ref. 33. The dashed line represents Eq. (20). The solid lines (drawn with same slope as the dotted line) are all guides to the eye.

To gain a survey regarding the correlation times of the various components in hydrated ethaline over a wide temperature range, in Fig. 8 we collect the timescales $\tau_J$ from rheology, $\tau_\varepsilon$ from dielectric spectroscopy, and $\tau_{T1}$ from NMR relaxometry. Since the $T_1$ times for $^2$H and for $^{17}$O are both governed by a first-order quadrupolar relaxation mechanism, considerations similar to those put forward in Sec IV A are not required. When analyzing the spin-lattice relaxation data we employed the procedure outlined in Ref. 40 which, in the fast motion regime, reduces to exploiting expressions like Eq. (12). As the mainframe of Fig. 8 shows, for hydrated ethaline the correlation times characterizing the different components follow the same overall temperature dependence. Our attempts to fit these data using a Vogel-Fulcher dependence[99] were not successful. However, we found that Waterton's law[100]

$$\tau = \tau_0 \exp\left[\frac{K}{T}\exp\left(\frac{C}{T}\right)\right], \quad (20)$$

which recently was demonstrated to work well for a wide range of glassformers,[101,102] provides an excellent description for hydrated ethaline. Unlike the Vogel-Fulcher approach, Eq. (20) does not imply a divergence of $\tau$ at a finite temperature. The parameters used for the fit to the $^2$H data of the EG-d$_4$:ChCl:H$_2^{17}$O sample that are displayed in Fig. 8 are $\tau_0 = 5 \times 10^{-13}$ s, $K = 420$ K, and $C = 361$ K. From Fig. 8 one recognizes that the fit captures also the $^{17}$O spin-relaxation based time scales very well. Furthermore, the extrapolation of the fit to low temperatures provides a good match with the dielectric and rheological data of hydrated ethaline.

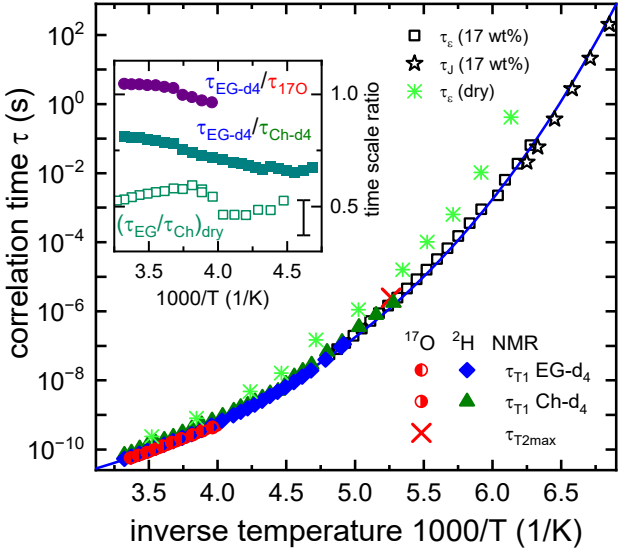

Fig. 8. Arrhenius plot including $^2$H and $^{17}$O spin relaxation based timescales for EG:ChCl-d$_4$:H$_2^{17}$O (▲ and ◐) as well as for EG-d$_4$:ChCl:H$_2^{17}$O (◆ and ◑). A timescale from the $^{17}$O $T_2$ maximum (×) is also shown. The solid line represents a fit using Eq. (20) to the data for the EG-d$_4$ component of hydrated ethaline. The dielectric time constant $\tau_\varepsilon$ and rheological one, $\tau_J$, both from the present work, refer to hydrated ethaline. For dry ethaline, dielectric relaxation times by Reuter et al.[31] (asterisks) are also shown. The inset summarizes various spin-relaxation based timescale ratios for dry[39] and for hydrated ethaline, with the error bar marking the typical experimental uncertainty.

From the mainframe of Fig. 8 it is obvious that, in harmony with the results from Fig. 7, also in the high-temperature region, $\tau_{T1,Ch}$ is larger than $\tau_{T1,EG}$. A more detailed comparison of the spin-relaxation based time scales is provided in the inset of Fig. 8 which shows $\tau_{T1,EG}/\tau_{T1,Ch}$ as a function of temperature. These results for hydrated ethaline are similar to the $(\tau_{T1,EG}/\tau_{T1,Ch})_{dry}$ ratio for dry ethaline.[39] This shows how well, in the water-in-DES regime, the H$_2$O additive blends into the native (dry) ethaline structure.

In the high-temperature regime (i.e., for $T > 220$ K) the dielectric time constants for dry ethaline were found compatible with $\tau_{EG}$ from NMR.[39] For hydrated ethaline, the dielectric radio-frequency measurements which are necessary to enable such a comparison are not available, so far. Close to the glass transition, Fig. 7 shows that the dielectric as well as the rheological time constants are longer than those from NMR. According to the arguments advanced in Sec IV A, this finding points to a scenario where small-angle jumps dominate the reorientation process in hydrated ethaline. For temperatures near $T_g$ a comparison with dry ethaline cannot be made, because its pronounced crystallization tendency precluded[39] the acquisition of stimulated-echo data.

Thus, one may say that in the particular case of hydrated ethaline it is the water component rather than (as usual) the salt components that acts as a low-temperature antifreeze. While water's performance is somewhat poor in that respect (cf. the enhanced crystallization tendency observed near 170 K for some hydrated samples), its beneficial effects are obviously sufficient to allow for NMR measurements continuously covering motional correlation times from shorter than nanoseconds to longer than milliseconds.

## V. Conclusion

By exploiting the unique capabilities of isotope edited NMR, the present work examined the component-selective dynamics of the different molecular moieties in ethaline that was hydrated to a water level of about 17 mol%. While the low-temperature $^2$H spectra of the EG and cholinium moieties are indicative for the usual quadrupolar parameters, those characterizing the $^{17}$O nucleus in H$_2$O are intermediate between the values known for crystalline and amorphous water on the one hand and those for liquid water and aqueous solutions on the other. Thus, $^{17}$O is particularly sensitive to the modulations of the intermolecular interactions that modify the electron distribution at the site of the oxygen probe.

In order to check for possible isotope effects with respect to the molecular time scales, we carried out dielectric and rheological experiments. We find that an $^{16}$O/$^{17}$O isotope effect is absent and, in harmony with previous work, a $^1$H/$^2$H isotope effect could also not be detected. Near $T_g$, it was found that the time scales from the dielectric susceptibility and from the shear compliance are somewhat longer than those from stimulated-echo spectroscopy, hinting at small-angle dominated reorientational motions that is typical for supercooled liquids. Furthermore, the timescale ratio from the stimulated-echo experiments reveal that the cholinium dynamics is about two times slower than that measured for EG. Conversely, within the discussed methodological uncertainties, rather similar underlying correlation times, $\tau_{H2O} \lesssim \tau_{EG}$, were inferred for the water and the EG molecules. Also at high temperatures, based on the spin-relaxation times measured for hydrated ethaline, it turned out that $\tau_{H2O} \approx \tau_{EG}$ and $\tau_{Ch} \approx 2\tau_{EG}$. The latter relation was similarly found for dry ethaline, and thus shows that water is well blended into the hydrogen-bond donor network in this DES.

## SUPPLEMENTARY MATERIAL

As supplementary material we document the absence of an $^{16}$O/$^{17}$O isotope effect, we provide low-temperature $^2$H and $^{17}$O spectra, show additional low-temperature $^2$H spin-lattice relaxation times, and present calculations to determine $\tau_{c,T2max}$ in the presence of a distribution of correlation times.

**ACKNOWLEDGEMENTS**

We thank Dr. Kevin Moch for helping with the rheological experiments. The Deutsche Forschungsgemeinschaft is thanked for supporting this work financially in the framework of project no. 444797029.

**DATA AVAILABILITY**

The data that support the findings of this study are available from the authors upon reasonable request.

Supplementary Material regarding the article

# Interplay of ethaline and water dynamics in a hydrated eutectic solvent: Deuteron and oxygen magnetic resonance studies of aqueous ethaline

Yannik Hinz and Roland Böhmer
*Fakultät Physik, Technische Universität Dortmund, D-44221 Dortmund, Germany*

Here we present

in Fig. S1 dielectric spectra to document the absence of an $^{16}O/^{17}O$ isotope effect,

in Fig. S2 a low-temperature $^2H$ spectrum,

in Fig. S3 a low-temperature $^{17}O$ spectrum including a numerical simulation,

in Fig. S4 additional low-temperature $^2H$ spin-lattice relaxation times, and in Fig. S5 calculations to determine $\tau_{c,T2\max}$ in the presence of a distribution of correlation times.

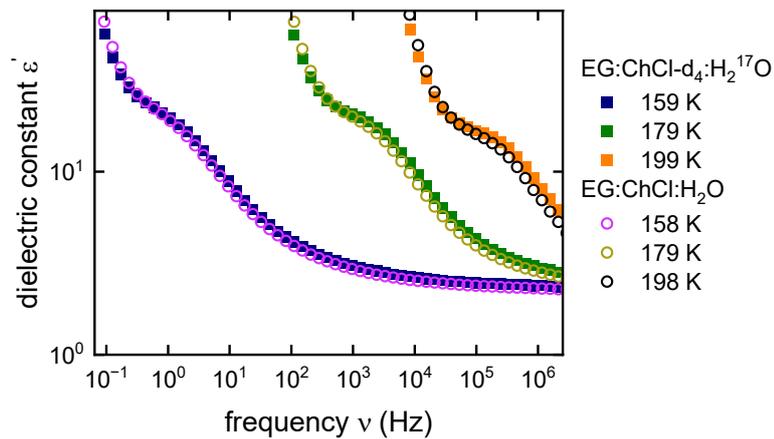

Fig. S1: Dielectric function $\varepsilon'(\nu)$ for EG:ChCl-d$_4$:H$_2^{17}$O and for EG:ChCl:H$_2$O at comparable temperatures. Clearly, a significant isotope effect is not discernible.

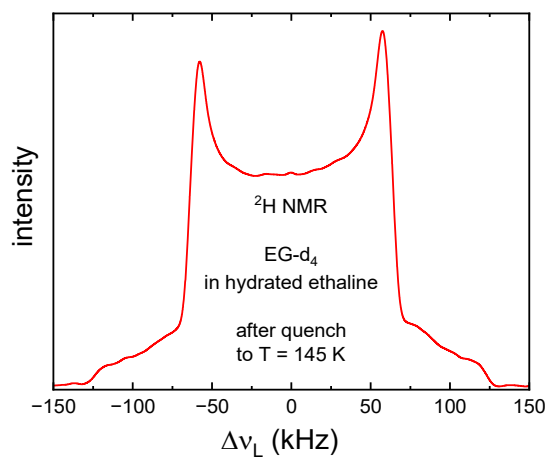

Fig. S2: Low-temperature $^2$H spectrum for EG-d$_4$:ChCl:H$_2^{17}$O recorded after a quench to 145 K. The pulse separation was 20 µs.

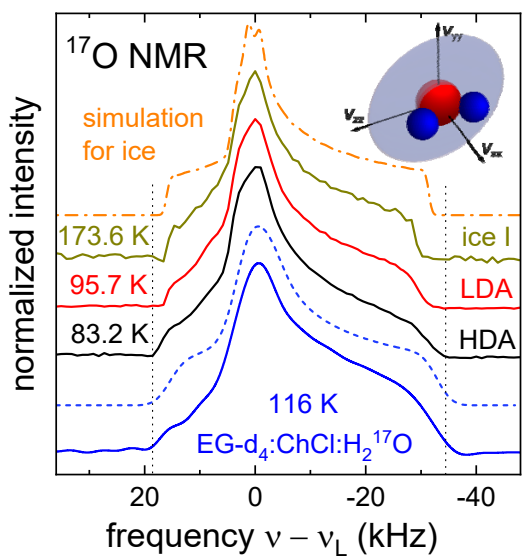

Fig. S3: Low-temperature $^{17}$O spectrum for EG-d$_4$:ChCl:H$_2^{17}$O. An explicit analysis shows that for this spectrum the central second moment is $\sqrt{M_{2c}} = 12.4$ kHz. A Gaussian apodization with a standard deviation of 500 Hz (full spectral width 1.2 kHz) was employed. The dashed line was calculated[S1] using a quadrupolar coupling constant $C_Q = 6.9$ MHz and an asymmetry parameter $\eta_O = 0.85$, corresponding to $P_O = 7.7$ MHz. The inset illustrates the orientation of the oxygen EFG tensor in the frame of the water molecule. The parameters chosen for the CSA tensor are as suggested in Ref. S2 for liquid water: $\Delta\sigma = 33$ ppm and $\eta_{CSA} = 0.03$. The apodization was taken to be 2 kHz as for the experimental spectrum. The orientation of the CSA with respect to the EFG tensor is assumed here as $\sigma_{xx} \parallel V_{zz}$, $\sigma_{yy} \parallel V_{xx}$, and $\sigma_{zz} \parallel V_{yy}$ so that the set of Euler angles is (90°,90°,0°). For comparison we show spectra measured for HDA, LDA, and ice I (all represented as solid lines) as well as a simulated ice spectrum ($C_Q = 6.66$ MHz and $\eta = 0.935$, dash-dotted line), all taken from Ref. S3. The vertical dotted lines aid in highlighting the slightly differing widths of the spectra.

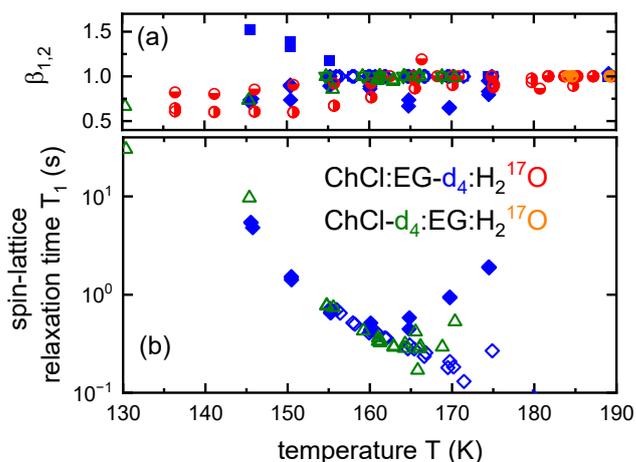

Fig. S4: (a) Kohlrausch exponents characterizing the longitudinal and the transverse spin relaxation times displayed in Fig. 3. Here the same symbols are used as in that figure. (b) Spin-lattice relaxation times measured in the low-temperature regime. Data recorded after slow cooling are represented by full symbols, those recorded after rapid cooling (quenching) by open symbols.

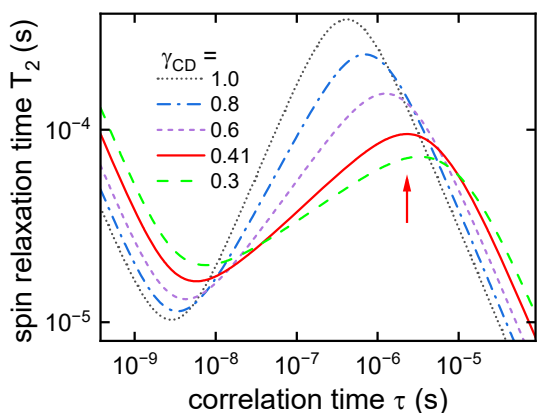

Fig. S5: Transverse dephasing times involving contributions from first- and second-order quadrupolar relaxation for a range of Cole-Davidson parameters $\gamma_{CD}$. The solid lines are calculated similar to the procedure employed in Ref. S4 (but adapted to $I = 5/2$) and Ref. S5 using $\omega_O = 2\pi \times 54.26$ MHz and a quadrupolar product $P_O = 8.5$ MHz. For the present analysis the curve for $\gamma_{CD} = 0.41$ is relevant. The time constant $\tau_{c,T2\max}$ is highlighted by the arrow.